\documentclass[%
 reprint,
 amsmath,amssymb,
 aps,
]{revtex4-1}

\usepackage{graphicx}
\usepackage{amssymb}
\usepackage{times}
\usepackage{amsmath}
\usepackage{amsthm}
\usepackage{array}
\usepackage{booktabs}
\usepackage{multirow}
\usepackage{setspace}
\usepackage{tabularx}
\usepackage{tabu}
\usepackage{siunitx}
\usepackage{subfigure}
\usepackage{extarrows}
\usepackage{hyperref}
\usepackage{lipsum}

\begin{document}

%\preprint{APS/123-QED}

\title{Mask-coding-assisted continuous-variable quantum direct communication with orbital angular momentum multiplexing}% Force line breaks with \\
%\thanks{A footnote to the article title}%

\author{Zhengwen Cao}
\author{Yujie Wang}
\author{Geng Chai}
\email{chai.geng@nwu.edu.cn}
\author{Xinlei Chen}
\author{Yuan Lu}
\affiliation{Institute for Quantum Information \& Technology (QIT), School 
of Information Science and Technology, Northwest University, Xi’an 
710127, China}%Lines break automatically or can be forced with \\%

\date{\today}% It is always \today, today,
             %  but any date may be explicitly specified

\begin{abstract}
Quantum secure direct communication (QSDC) is a approach of communication to transmit secret messages base on quantum mechanics. Different from the quantum key distribution, secret messages can be transmitted directly on quantum channel with QSDC. Higher channel capacity and noise suppression capabilities are key to achieving long-distance quantum communication. Here we report a continuous-variable QSDC scheme based on mask-coding and orbital angular momentum, in which the mask-coding is employed to protect the security of the transmitting messages and to suppression the influence of excess noise. And the combination of orbital angular momentum and the information block transmission can effectively improve the secrecy capacity. In the $800$ information blocks $\times$ 1310 bits length 10-km experiment, the results show that the statistical average of bit error rate has decreased by 36$\%$ and reached 0.38$\%$ comparing to existing solutions, and the system excess noise is 0.0184 SNU, and the final secrecy capacity is 6.319$\times10^{6}$ bps. Therefore, this scheme reduces error bits while increasing secrecy capacity, providing a solution for long-distance large-scale quantum communication.
\end{abstract}

\pacs{Valid PACS appear here}% PACS, the Physics and Astronomy
                             % Classification Scheme.
%\keywords{Suggested keywords}%Use showkeys class option if keyword
                              %display desired
\maketitle

%\tableofcontents
\section{\label{sec:level1}Introduction}
With the continuous development of science and technology, information security has gradually become the focus of attention. And the security of communication systems is seriously threatened by quantum computing and distributed computing. In order to combat the upheaval of the information age and ensure the security of information transmission, communication approaches based on the basic principles of quantum mechanics came into being. With the rapid rise of quantum communication, it has brought more possibilities to our lives \cite{r1}. At present, quantum communication mainly includes quantum key distribution (QKD) \cite{r23, r26, r29,r24,r33} %%quantum teleportation \cite{r24}
 and quantum secure direct communication (QSDC) \cite{r5}.

Unlike QKD, QSDC was proposed in 2000, which allows direct transmission of secret messages rather than keys \cite{r5}, and reduces steps in the communication process and the complexity of information transmission while improving efficiency \cite{r2}. In early research on QSDC, many relevant protocols \cite{r3, r4,r6} were developed for different information carriers. %, such as entangled states. %In 2001,  deterministic secure communication was clearly defined by Beign et al. \cite{r3, r4}, upon which Boström and Felbinger proposed Ping-Pong scheme \cite{r4}. In their protocol, message mode or communication mode was selected randomly by both sides of the communication for information exchange. However, the Ping-Pong scheme \cite{r4} had been proved that it had unsafe factors \cite{r6}. 
Two-step QSDC protocol with entangled state was proposed by Deng $et.\ al.$ in 2003 \cite{r7}, whose security was verified by comparing the joint statistical results of the first security judgment with the threshold value. Moreover, DL04 protocol with single photon was proposed by Long $et.\ al.$ \cite{r8}, promoting the development of QSDC. Them have been proved to be information theoretic security and theoretical feasible. In addition, industry scholars have also proposed various QSDC protocols, such as  entangled states protocol \cite{r40,r41}, and Greenberger-Horne-Zeilinger states protocol \cite{r42,r43},  quantum-memory-free QSDC protocol \cite{r44,r45}, device-independent QSDC,  measurement device-independent QSDC protocol \cite{r35,r36,r37,r38,r39}.
 
In recent years, quantum direct communication technology has also been experimentally validated, such as entangled states and single photon QSDC experiments \cite{r28,r30}, whose have been reported for verifying the principles of QSDC protocol based on single photon and entanglement. The first QSDC experiment based on long-range entanglement was reported, and the performance of QSDC system was theoretically analyzed, and the results shown that QSDC can be achieved on several hundreds of kilometers of fiber optic links under current optical communication technology, providing a method to extend QSDC to further distances \cite{r10}. Long $et.\ al.$ reported an experimental implementation of a practical quantum secure communication system using the DL04 protocol \cite{r11}. In order to push QSDC into practical applications, some critical issues must be addressed. Cao $et.\ al.$ reported the continuous variable QSDC experimental demonstration, verifying the feasibility and effectiveness of the communication method in fiber optic channels, and proposed a parameter estimation for signal classification in actual channels, laying the foundation for grading reconciliation \cite{r9}. 

Although QSDC has made significant progress in both theory and engineering, it still needs to be further promoted in terms of engineering implementation and practicality. In order to achieve long-distance quantum communication, communication protocols need to have higher channel capacity and noise suppression capabilities. The multiplexing of orbital angular momentum (OAM) and mask coding is an effective way to solve this problem. OAM is a basic photon physical quantity, which has great application potential in optical research \cite{r31}. Photonic OAM has good characteristics to realize multiplexing technology. At present, OAM is often employed in the following aspects: using higher dimensional Hilbert space to encode information in quantum information processing systems \cite{r32}, and used as the efficient carrier of information reuse. The OAM state has infinity and orthogonality, which is an effective way to implement multiplexing technology and can effectively increase the capacity of the QSDC protocol. Additionally, incorporating mask coding technology in the protocol can effectively enhance the anti-interference ability of information during transmission \cite{r25,r34,r12}.

Here we firstly propose a mask-coding-assisted continuous-variable quantum direct communication based on OAM multiplexing, and verified its feasibility through experiments. According to Wyner’s wiretap channel model, the performance of the proposed protocol has been analyzed. Then, in our experiment, the Hash function is use to mask the secret messages into random sequence for improving the security of information transmission. Taking advantage of the orthogonality between OAM modes, each mode can be regarded as an independent sub-channel, which changes the information block from serial transmission to parallel transmission and improves the communication capacity of the system. Finally, in the $800$ information blocks $\times$ 1310 bits length 10-km experiment, the results show that the statistical average of bit error rate has decreased by 36$\%$ and reached 0.38$\%$ comparing to existing solutions, and the system excess noise is 0.0184 SNU, and the final secrecy capacity is 6.319$\times10^{6}$ bps, fully validating the feasibility of the proposed protocol.

 The paper is organized as follows. In the Sec. \ref{sec:level2}, we provided a detailed introduction to the designed protocol and analyzed its performance based on the Wyner's model; in the Sec. \ref{sec:level3}, the experimental results are analyzed based on parameter estimation by building an experimental platform. The conclusion and discussion are presented in Sec. \ref{sec:level4}.
\section{\label{sec:level2}Protocol and performance analysis}
\begin{figure*}
   \centering
   \includegraphics[scale=0.9]{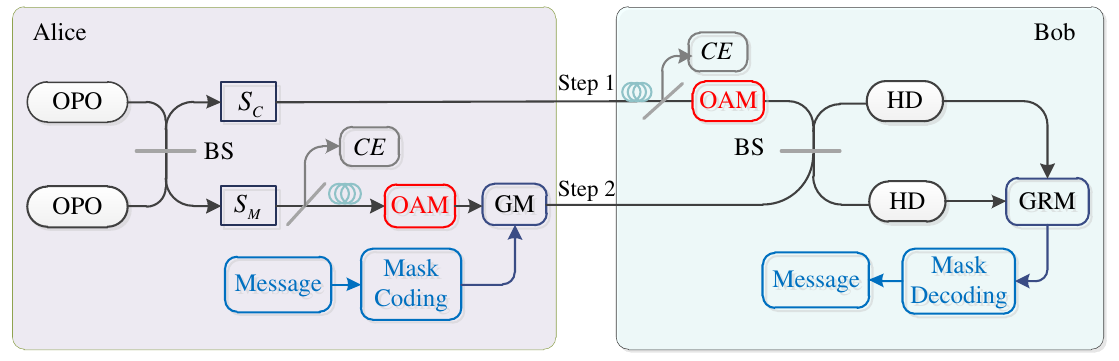}
   \caption{System diagram of Mask-coding-assisted CV-QSDC based on OAM multiplexing. $OPO$: optical parametric oscillator; $BS$: beam splitter; $S_C$: detection light; $S_M$: signal light; $CE$: checking eavesdropping; $OAM$: orbital angular momentum; $GM$: Gaussian mapping; $HD$: homodyne detection; $GRM$: Gaussian remapping. }
   \label{fig1}
 \end{figure*}

The first part of this section introduces the content of the protocol as shown in Fig. \ref{fig1}. And combining with the Wyner's eavesdropping channel model as shown in Fig. \ref{fig2}, the second part specifically analyze the improvement, brought by the OAM multiplexing and mask coding of this protocol.
\subsection{\label{sec:level21}Protocol description}
{\bf{Step 1 (Preparation of entangled light)}}:
Alice uses a 50:50 beam splitter to couple two single-mode squeezed entangled states to prepare two-mode squeezed entangled state, which respectively recorded as $S_C$  (detection light) and $S_M$  (signal light). Alice retains the signal light $S_M$   and sends the detection light  $S_C$ after OAM multiplexing to Bob through the quantum channel. Specifically, the quadrature components of two single-mode squeezed states are expressed as:
\begin{eqnarray}
\label{eq1}
\left\{
\begin{aligned}
x_1 &= \exp(-r)\left | 0 \right \rangle_{x_1},\ p_1=\exp(r)\left | 0 \right \rangle _{p_1},\\
x_2 &= \exp(-r)\left | 0 \right \rangle _{x_2},\ p_2=\exp(r)\left | 0 \right \rangle _{p_2},
\end{aligned}
\right.
\end{eqnarray}
where $r(>0)$ is the squeezing factor, $x$ and $p$ denote the quadrature position and quadrature momentum, respectively. $\left | 0 \right \rangle$ denotes the vacuum state. Then the quadrature components of the two-mode squeezed 
entangled state can be expressed as:
\begin{eqnarray}
\label{eq2}
\left\{
\begin{aligned}
x_{S_C} &= {\frac{1} {\sqrt{2}}}(x_1+x_2),\ p_{S_C} = 
{\frac{1} {\sqrt{2}}}(p_1+p_2),\cr
x_{S_M} &= {\frac{1} {\sqrt{2}}}(x_2-x_1),\ p_{S_M} = {\frac{1} 
{\sqrt{2}}}(p_2-p_1).
\end{aligned}
\right.
 \end{eqnarray}
 
\begin{figure*}
   \centering
   \includegraphics[scale=0.3]{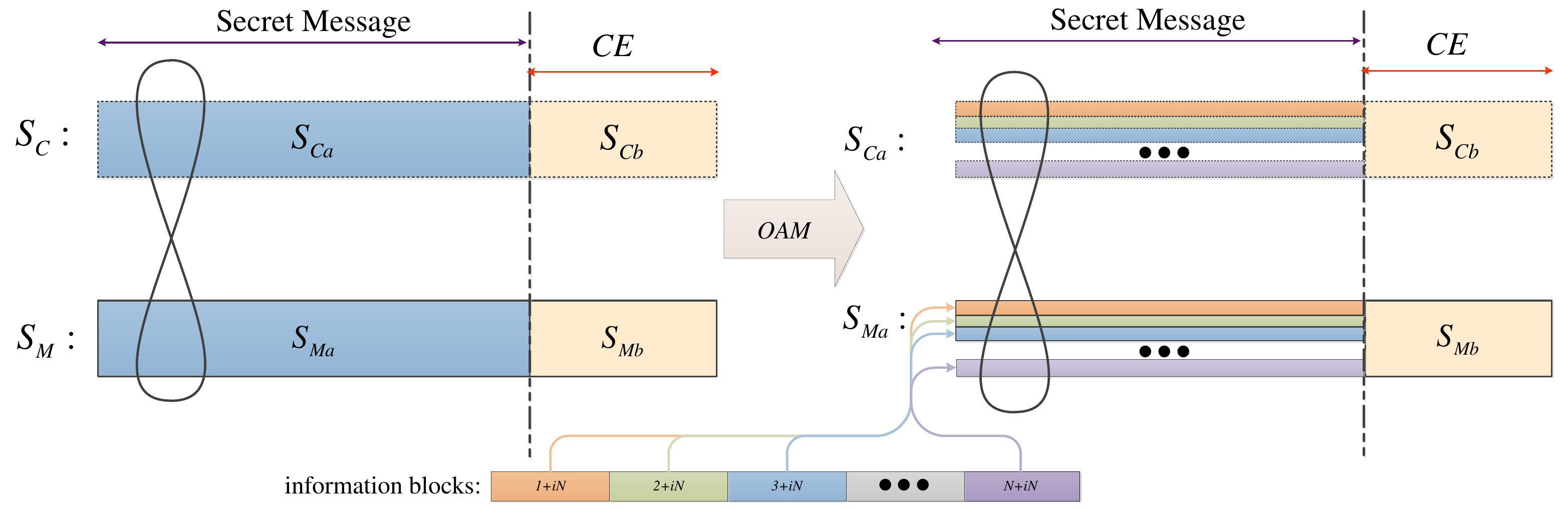}
   \caption{Preprocessing of entangled light. $S_C$: detection light; $S_M$: signal 
   light; the subscript $a$ is used for security detection and identity authentication, while the subscript $b$ is used for information carrying and measurement; $CE$: checking eavesdropping; $OAM$: orbital angular momentum; $N$: $OAM$ multiplexing modes. }
   \label{fig00}
 \end{figure*}

 {\bf{Step 2 (Channel security detection)}}: Alice divides the detection light $S_C$ and signal light $S_M$ into two parts, $a$ and $b$, respectively. The preprocessing scheme for entangled light is shown in Fig. \ref{fig00}. Furthermore, $b$ is used for security detection and identity authentication, while $a$ is used for information carrying and measurement. In order to achieve OAM multiplexing of beams, the $S_\text{{Ca}}$  is transformed by a group of spatial light modulators (SLMs) with spiral phase of $\exp (il_p\theta )$ to obtain $S_\text{{Cai}}$. And Alice performs the same operation on $S_\text{{Ma}}$ after channel security detection, identity authentication, and mask coding:
\begin{eqnarray}
E_{S_\text{{Cai}}}^\text{{MUX}} (\boldsymbol {r},\theta ,t)=\sum_{p=1}^{N} E_{S_\text{{Cai}}} \left ( \boldsymbol {r},t \right ) \exp(il_p\theta ),
\label{eq9} 
\end{eqnarray}
where, $E_{S_\text{{Cai}}}^\text{{MUX}} \left (\boldsymbol {r}, \theta, t \right ) $ is the quantized light field of $S_\text{{Cai}}$,  and $E_{S_\text{{Cai}}} \left (\boldsymbol {r}, t \right ) $ is the quantized light field of $S_\text{{Cai}}$.The definition of $E_{S_\text{{Cai}}} \left (\boldsymbol {r}, t \right ) $ will be provided in the following text. Alice randomly selects time slots to measure the quadrature components of $S_\text{{Mb}}$, and then announces Bob the selected time slots and measurement results. After receiving the detection light $S_C$, Bob selects the same time slots as Alice to measure the quadrature components of $S_\text{{Cb}}$, and makes a joint judgment on  the measurement results between Alice and Bob: 
\begin{eqnarray}
\left \langle \left [ \bigtriangleup \left ( \frac{x_{mA}\pm 
x_{mB}}{\sqrt{2} }  \right )  \right ] ^{2}  \right \rangle +\left 
\langle \left [ \bigtriangleup \left ( \frac{p_{mA}\pm p_{mB}}{\sqrt{2} 
}  \right )  \right ] ^{2}  \right \rangle < 2 ,
\label{eq3}
\end{eqnarray}
where, $x(p)_{mA}$ and $x(p)_{mB}$  represent the measurement results of both communication parties. If Bob's measurement results and Alice's measurement results meet the Eq. (\ref{eq3}), the quantum channel is safe and communication continues; otherwise, the communication is terminated.

{\bf{Step 3 (Identity authentication)}}: Similarly, Bob randomly selects other time slots on the remaining $S_\text{{Cb}}$ for measurement, and then sends the time slots and measurement results to Alice. Alice also conducts joint measurement through Eq. (\ref{eq3}). The purpose of this phase is for Alice to verify and confirm Bob's identity. If the identity is legal, Alice go to the next step. Otherwise, the communication is terminated.

{\bf{Step 4 (Information mask coding)}}: Alice employs mask coding to mask the secret messages to obtain a random message sequence $M$ \cite{r15}. This protocol takes the Hash function cluster designed by Toeplitz matrix as an example to illustrate the mask coding:

\begin{itemize}
\item 
 Alice randomly selects a random code sequence $s$ of length $k+n-1$ and constructs a $k\times n$ binary Toeplitz matrix(where $k$ is the length of the secret messages and $n$ is the length of the masked target). Toeplitz matrix, also known as constant quantity matrix, has equal elements on the same diagonal parallel to the main diagonal, each row of the matrix is the result of moving one element to the right relative to the row above it, and the newly added position on the left is filled by another sequence. The $k \times n$ Toeplitz matrix can be represented as follows \cite{r12}:
\begin{equation}
\label{eq4}
\left[
\begin{array}{cccccc}
a_{1,1} & a_{1,2} & a_{1,3} & a_{1,4}& \cdots &a_{1,n} \\
a_{1,n} & a_{1,1} & a_{1,2}& a_{1,3}& \cdots &a_{1,n-1}\\
a_{1,n-1} & a_{1,n} & a_{1,1}& a_{1,2}& \cdots &a_{1,n-2}\\
\vdots  & \vdots  & \vdots & \vdots & \ddots  &\vdots \\
a_{1,k} & a_{1,k-1} & a_{1,k-2} & a_{1,k-3}& \cdots &a_{1,n-k+1}
\end{array}
\right]_{k \times n} .     
\end{equation}

Taking secret message $m=001,n=5$ as an example, Alice selects a 
random number sequence $s=a_{1}a_{2}a_{3}a_{4}a_{5}a_{6}a_{7}=1100100$ and constructs a Toeplitz matrix 
as shown in Eq. (\ref{eq5});
\begin{equation}
\begin{split}
T_{oe}=\left[
\begin{array}{ccccc}
a_{1} & a_{2} & a_{3} & a_{4} & a_{5}\\
a_{7} & a_{1} & a_{2} & a_{3} & a_{4}\\
a_{6} & a_{7} & a_{1} & a_{2} & a_{3}\\
\end{array}
\right]_{3\times5} 
=\left[
\begin{array}{ccccc}
1 & 1 & 0 &0 & 1\\
0 & 1 & 1 &0 & 0\\
0 & 0 & 1 &1 & 0\\
\end{array}
\right] .
\label{eq5} 
\end{split}   
\end{equation}
\end{itemize}

\begin{itemize}
\item
 Alice transforms the Toeplitz matrix in rows to get standardized matrix;
\begin{equation}
T_{oe}'=[A_{T_{oe}}|E]=\left[
\begin{array}{cc|ccc}
0 & 1 & 1 &0 & 0\\
0 & 1 & 0 &1 & 0\\
1 & 1 & 0 &0 & 1\\
\end{array}
\right]   .
\label{eq6}   
\end{equation}

Among them, the sub matrix $A_{T_{oe}}$ and the  randomized redundant sequence $R_{oe}=[1\ 0]^{\text{T}}$  are used to construct the matrix $[R_{oe}\ \  m-A_{T_{oe}}R_{oe}]^{\text{T}}$ , as shown in Eq. (\ref{eq7}), and then the masked message sequence $M=10000$ is obtained,
\begin{equation}
\begin{split}
M&=\left[
\begin{array}{ccccc}
R_{oe}\\
 m-A_{T_{oe}}R_{oe}\\
\end{array}
\right]             
\\
&=\left[
\begin{array}{ccccc}
{
\left[
\begin{array}{cc}
1\\
0\\
\end{array}
\right]}\\
\\
{
\left[
\begin{array}{ccc}
0\\
0\\
1\\
\end{array}
\right]}-
{
\left[
\begin{array}{ccc}
0&1\\
0&1\\
1&1\\
\end{array}
\right]}
{
\left[
\begin{array}{cc}
1\\
0\\
\end{array}
\right]}
\\
\end{array}
\right] 
=\left[
\begin{array}{ccccc}
1\\
0\\
0\\
0\\
0\\
\end{array}
\right]  .
\label{eq7}   
\end{split}  
\end{equation}
\end{itemize}

\begin{itemize}
\item
Alice sends Bob the starting digit of the random 
sequence $s$ used through the classic authentic channel after Demultiplexing and Measurement, so that Alice and Bob share the same Hash function to assist Bob extract secret information.
\end{itemize}

{\bf{Step 5 (Gaussian mapping):}} Alice first unifies the mask message sequence $M$ to obtain a balanced sequence ${M}' $, so that the sequence ${M}' $ has a stable uniform distribution \cite{r18}. Then the balanced sequence is mapped to a set of Gaussian random numbers with zero mean and variance $V_a$. %The corresponding relationship between the Gaussian random number and the information block is shown in Fig. \ref{fig1}. 
This protocol takes each information block containing 3 bits as an example \cite{r13}, and the Gaussian random number is divided into 8 intervals according to the principle of equal probability. The corresponding Gaussian intervals are,  $\left [   -7 ,-3.25\right ], \left ( -3.25,-1.91\right ], 
\left ( -1.91,-0.90\right ], \left ( -0.90,0\right ], \\ \left ( 0,0.90\right ],
\left ( 0.90,1.91\right ], \left ( 1.91,3.25\right ],  \left ( 3.25,+  7  \right ] $, and the eight intervals from left to right correspond to 000 to 111 of 3-bit binary string.
 
{\bf{Step 6 (Information multiplexing):}} Alice generates OAM signal lights with different modes $\left \{ S_{\text{Mai}}|i = 1,2,\dots,N\right \} $ from signal light $S_\text{{Ma}}$ through spatial light modulators \cite{r14}, and groups the information blocks according to OAM's  modes $N$. Each information block is modulated on the quadrature component ($x$ or $p$) of $S_\text{{Mai}}$ in order to obtain the final information beam $S_{\text{info}}$, as shown in the Fig. \ref{fig00}. Specifically, the quantized light field a of the signal light $S_{\text{Mai}}$  can be expressed as:
\begin{eqnarray}
E_{S_\text{{Mai}}} \left ( \boldsymbol {r},t \right ) = K\times i\left [ ae^{-i \left (\omega 
t-\boldsymbol {k}\cdot\boldsymbol {r}\right) } -a^{\dagger}e^{i(\omega 
t-\boldsymbol {k}\cdot\boldsymbol {r})} \right ]  
\label{eq8}, 
\end{eqnarray}
where $K=\left ( \frac{\hbar \omega }{2\varepsilon _0V}  \right )^{\frac{1}{2} }$, $\hbar $ is the Planck constant. $\varepsilon _0$  is the dielectric constant, $\omega  $  is the angular frequency,  $\boldsymbol {k}$ represents the wave number, $\boldsymbol {r}$ represents the wave vector.  $a$ and $a^{\dagger}$ represents the generation operator and the annihilation operator, respectively. The quantized light field of the OAM beam generated by Alice is represented as:
\begin{eqnarray}
E_{S_\text{{Mai}}}^{\text{MUX}} (\boldsymbol {r},\theta ,t)=\sum_{p=1}^{N} E_{S_\text{{Mai}}} \left ( \boldsymbol {r},t \right ) \times 
\exp(il_p\theta ),
\label{eq9} 
\end{eqnarray}
where $l_p$ represents the topological charge of OAM beam, and $N$ is the number of modes used for multiplexing. And the quantized light field of the final signal light beam $S_{\text{info}}$  generated by Alice is represented as:
\begin{widetext}
\begin{equation}
\begin{split}
E_{S_{\text{info}}}(\boldsymbol{r},\theta ,t)&=  E_{S_\text{{mi}}}^\text{{MUX}} (\boldsymbol{r},\theta 
,t)\exp\left ( i\frac{\pi V_b }{V_\pi }  \right )
\times\exp\left [ 
i\sum_{p=1}^{N}\frac{\pi A_p}{V_{\pi}}\cos\left ( \omega _kt+\phi _p 
\right )    \right ]  \\
&=E^{A_p}\left \{ \hat{a}_A \exp\left [ i\left (  
\boldsymbol{k}\cdot\boldsymbol{r}-\omega _kt\right )  \right ] 
+\hat{a}_A^{\dagger}
\exp\left [ -i\left ( 
\boldsymbol{k}\cdot\boldsymbol{r}-\omega _kt \right )  \right ]    
\right \} ,
\label{eq99} 
\end{split}
\end{equation}
\end{widetext}
where $V_b$ represents bias voltage of amplitude modulator, $V_{\pi }$ represents half wave voltage of phase modulator, $A_{p}$ represents a 
random number obeying Rayleigh distribution for amplitude modulation, 
$\phi _p$ represents a uniformly distributed random number for phase 
modulation, $\hat{a}_A$ represents the dimensionless complex amplitude 
operator after Gaussian modulation.

{\bf{Step 7 (Demultiplexing and Measurement):}}  After receiving the signal light ${S}'_{\text{info}}$  (after channel transmission), Bob separates $N$ OAM beams of ${S}' _{\text{Cai}}$ from the detection light ${S}'_{C}$   (after channel transmission), also separates the $N$ OAM beams signal light ${S}'_{\text{Mai}}$  from ${S}'_{\text{info}}$. The detection light ${S}'_{\text{Ci}}$ with the same mode OAM signal light ${S}'_{\text{Mi}}$ can be measured and recovered to Gaussian plane light from the spiral phase. After passing through the quantum channel, the optical light of the detection light and signal light received by Bob are:
 
 \begin{eqnarray}
\left\{
\begin{aligned}
&E_{S'_C}^{Bob}(t)= E_{S_C}\cdot \exp (i\phi ),
\label{eq10}\cr
&E_{S'_\text{{info}}}^{Bob}(t)= E_{S_\text{{info}}}\cdot \exp (i\phi ),
\label{eq10}
\end{aligned}
\right.
\end{eqnarray}
 where $\exp (i\phi )$ represents the evolution of channel transmission \cite{r14}. Bob performs CV Bell detection on the detection light and signal light, and the measurement results can be written as:
\begin{eqnarray}
I^{l_q} \propto \int_{0}^{+  \infty } \int_{0}^{2\pi } \left | 
E_{S'_\text{{Mai}}}^\text{{MUX}} + \left[E_{S'_\text{{Cai}}}^\text{{MUX}} \right ]^{\ast }\right | 
^2\boldsymbol{r}d\boldsymbol{r}d\theta ,
\label{eq12}
\end{eqnarray}
where $*$  represents conjugate transformation. The result after 
removing the direct-current component can be expressed as:
\begin{widetext}
\begin{equation}
\begin{split}
I^{l_p}&=Re\left \{ \int_{0}^{+  \infty }\int_{0}^{2\pi } 
E_{S'_\text{{Mai}}}^\text{{MUX}}+\left [ E_{S'_\text{{Cai}}}^\text{{MUX}}  \right ]^\ast  
\boldsymbol{r}d\boldsymbol{r}d\theta     \right \} \cr
&=Re\left \{ \int_{0}^{+  \infty 
}\int_{0}^{2\pi } \left[\sum_{p=1}^{N} E_{S'_\text{{Mai(p)}}} 
\exp(il_q\theta )\right]  
 E_{S'_\text{{Cai}}} \exp(-il_q\theta )  
\boldsymbol{r}d\boldsymbol{r}d\theta    \right  \} .
\label{eq13}
\end{split}
\end{equation}
\end{widetext}

In addition, different modes of OAM light satisfy orthogonal 
characteristics, and only vortex light with the same mode can be 
successfully demultiplexed and measured during detection. The joint 
measurement result is represented as:
\begin{eqnarray}
\label{eq14}
\left\{
\begin{aligned}
x_m&=\frac{1}{\sqrt{2} } \left ( x'_{S'_\text{{Mai}}} -x'_{S'_\text{{Cai}}}\right ), \cr
p_m&=\frac{1}{\sqrt{2} } \left ( p'_{S'_\text{{Mai}}} +  p'_{S'_\text{{Cai}}}\right ) .
\end{aligned}
\right.
\end{eqnarray}
 
{\bf{Step 8 (Information extraction):}} Bob then remaps the measured results through information reconciliation to obtain the masked message sequence. Alice sends the description sequence $s$ of the Hash function to Bob. Subsequently, Bob uses sequence $s$ to perform mask decoding on the masked message sequence. Corresponding to \textbf{Step 4}, the original secret message $m=001$ can be obtained. At this point, the communication between the two parties ends.
\begin{equation}
\begin{split}
m&=T_{oe}'M\\
&=\left[
\begin{array}{ccccc}
0 & 1 & 1 &0 & 0\\
0 & 1 & 0 &1 & 0\\
1 & 1 & 0 &0 & 1\\
\end{array}
\right]     
 \left[
\begin{array}{ccccc}
1\\
0\\
0\\
0\\
0\\
\end{array}
\right]=
 \left[
\begin{array}{ccc}
0\\
0\\
1\\
\end{array}
\right] .
\end{split}
\label{eq15}
\end{equation}
 
 \subsection{\label{sec:level22}System performance analysis}
\begin{figure}
   \centering
   \includegraphics[scale=0.8]{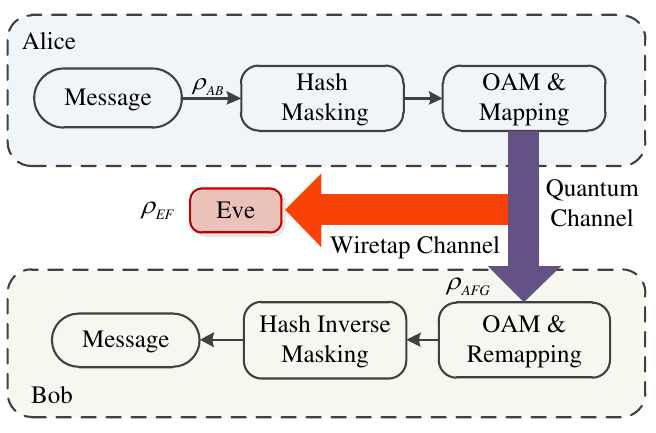}
   \caption{Wyner's wiretap channel model. Quantum Channel: main channel between Alice and Bob; Wiretap Channel:  eavesdropping channel of Eve; $OAM$: orbital angular momentum; $\rho_\text{AB}$: initial state between Alice and Bob; $\rho_\text{EF}$: auxiliary state of Eve; $\rho_\text{AFG}$: receiving state after eavesdropping and interference. }
   \label{fig2}
 \end{figure}  
 
 The channel with eavesdroppers is called eavesdropping channel, and the secrecy capacity is the maximum secure transmission rate between Alice and Bob. According to the communication model in the Wyner’s wiretap channel theory \cite{r16}, as shown in Fig. \ref{fig2}, the difference between the main channel capacity and the eavesdropping channel capacity in the Wyner’s wiretap channel model is defined as the secrecy capacity between the two communication parties:
\begin{eqnarray}
\begin{split}
C_s&\ge Q_B[2-h_4(e)]-Q_E\left[h\left ( \epsilon _x \right )+ h\left ( \epsilon _z \right )\right] \\
&\overset{\text{def}}{\longrightarrow}C_M-C_W= Q_BI_{AB}-Q_E\chi _\text{BE},
\end{split}
\label{eq16}
\end{eqnarray}
where $C_S$ represents the secrecy capacity, $C_M$ represents the main channel capacity, $C_W$ represents the eavesdropping channel capacity, $I_\text{{AB}}$ represents the amount of mutual information between Alice and Bob \cite{r17}, and $\chi _\text{{BE}} $ represents the Holevo bound that eavesdropper Eve can obtain. Considering the block reception rate of Bob $Q_B$ and Eve $Q_E$ that represent the receiving capability of Bob and Eve channels. The secrecy capacity with orbital angular momentum multiplexing can be calculated:
\begin{eqnarray}
\begin{split}
C_{s}^\text{{MUX}}&=N\times(C_M-C_W)\cr
&=N\times\left(Q_BI_\text{{AB}}-Q_E\chi 
_\text{{BE}}\right),
\label{eq18}
\end{split}
\end{eqnarray}
where $N$ is the number of OAM modes. Taking the measurement of quantum signals using CV Bell detection as an example,  $I_\text{{AB}}$ can be represented as:
\begin{eqnarray}
I_{AB}=2\times \frac{1}{2} \log_2\left(\frac{V_B}{V_{B|A}} 
\right)=\log_2\left(\frac{V+\chi _\text{{tot}}}{1+\chi _\text{{tot}}} \right),
\label{eq17}
\end{eqnarray}
where ${V=V_a+1}$, $\chi _\text{{tot}}=\chi _\text{{line}}+\frac{\chi_h}{T}$ represents the total noise, $\chi _{line}=1/T-1+\varepsilon $ represents the linear noise present in the channel, $\chi _{h}=2(1+v_\text{{el}})/\eta -1$  represents the detection noise from CV Bell detection , wherein $v_{el}$ represents the electrical noise and $\eta$ represents the detection efficiency.  Moreover $\chi _\text{{BE}}$ can be represented as:
\begin{eqnarray}
\chi _\text{{BE}}=\sum_{i=1}^{2} G\left ( \frac{\lambda _i-1 }{2} \right ) 
-\sum_{i=3}^{5} G\left ( \frac{\lambda _i-1 }{2} \right ), 
\label{eq21}
\end{eqnarray}
where, $ G(x)=(x+1)\log_2(x+1)-x\log_2(x)$ . $\lambda _i$  is the symplectic eigenvalue calculated according to the covariance matrix corresponding to the quantum state $\rho_\text{{AB}}$, $\rho_\text{{EF}}$, $\rho_\text{{AFG}}$ at each stage, which can be obtained from Eq. (\ref{eq24}):
\begin{eqnarray} 
\begin{split}
&\left\{
\begin{aligned}
\lambda _{1,2}&=\frac{1}{2} \left [ A\pm \sqrt{A^2-4B}  \right ] ,\cr
\lambda _{3,4}&=\frac{1}{2} \left [ C\pm \sqrt{C^2-4D}  \right ] ,\cr
\lambda _{5}&=1 ,
\end{aligned}
\right.
\label{eq24}
\end{split}
\end{eqnarray}
where,

\begin{widetext}
\begin{equation}
\begin{split}
\label{eq26}
&\left\{
\begin{aligned}
A&=V^2(1-2T)+2T+T^2(V+\chi _\text{{line}}),\cr
B&=T^2(V\chi _\text{{line}}+1)^2,\cr
C&=\frac{A\chi _\text{{het}}^{2}+B+1+2\chi _\text{{het}}(V\sqrt{B}+T(V+\chi _\text{{line}}) 
)+2T(V^2-1) }{T^2(V+\chi _\text{{tot}})^2} ,\cr
D&=\frac{(V+\sqrt{B} \chi _\text{{het}})^2}{T^2(V+\chi _\text{{tot}})^2} 
,
\end{aligned}
\right.
\end{split}
\end{equation}
\end{widetext}
according to the calculation process of $G(x)$, there is $G\left ( 
\frac{\lambda _5-1}{2}  \right )=0 $.

Then we can get the system performance of Gaussian mapping CV-QSDC based on OAM and mask coding when the eavesdropper attacks the communication system with collective attack. Four modes of OAM beams are selected in this protocol, and the secrecy capacity can be expressed as:
\begin{equation}
C_s^\text{{MUX}}=4\times \left ( Q_BI_\text{{AB}}-Q_E\chi _\text{{BE}} \right ).
\label{eq27}
\end{equation}

\begin{figure}
   \centering
   \includegraphics[scale=0.6]{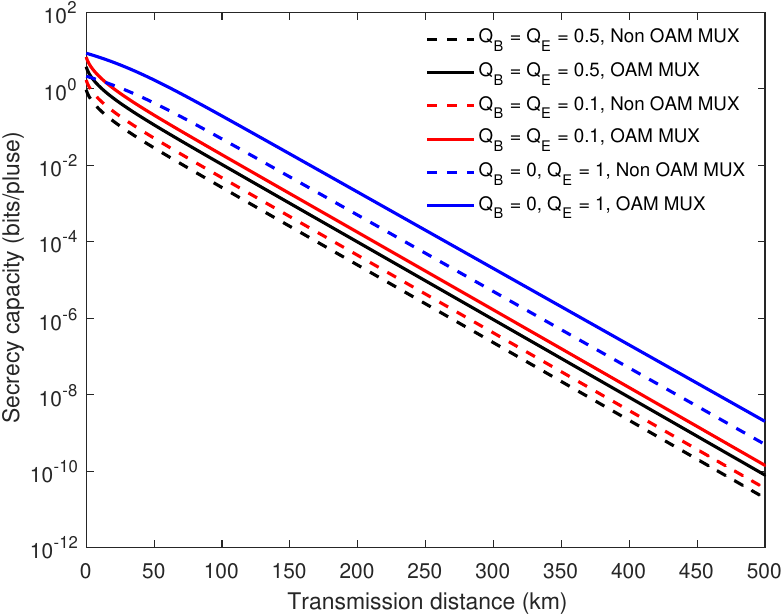}
   \caption{Secrecy capacity. The dotted line indicates the secrecy 
   capacity without orbital angular momentum multiplexing under 
   different $Q_B$ and $Q_E$, and the solid line indicates the secrecy 
   capacity with orbital angular momentum multiplexing under different 
   $Q_B$ and $Q_E$.}
   \label{fig3}
 \end{figure}

In combination with Sec.  \ref{sec:level22}, the relationship between secrecy capacity and transmission distance is calculated for different $Q_B$ and $Q_E$. According to Fig. \ref{fig3}, in different transmission environments, this scheme can improve system performance compared with existing schemes. On the other hands, under the same multiplexing conditions, the lower the system error rate, the more significant the performance improvement effect of this scheme. Furthermore, in this scheme, a mask coding is used to protect the transmission security of massages and resist the influence of channel noise. The combination of orbital angular momentum and information block transmission can effectively improve system performance.

  \begin{figure*}
   \centering
   \includegraphics[scale=0.60]{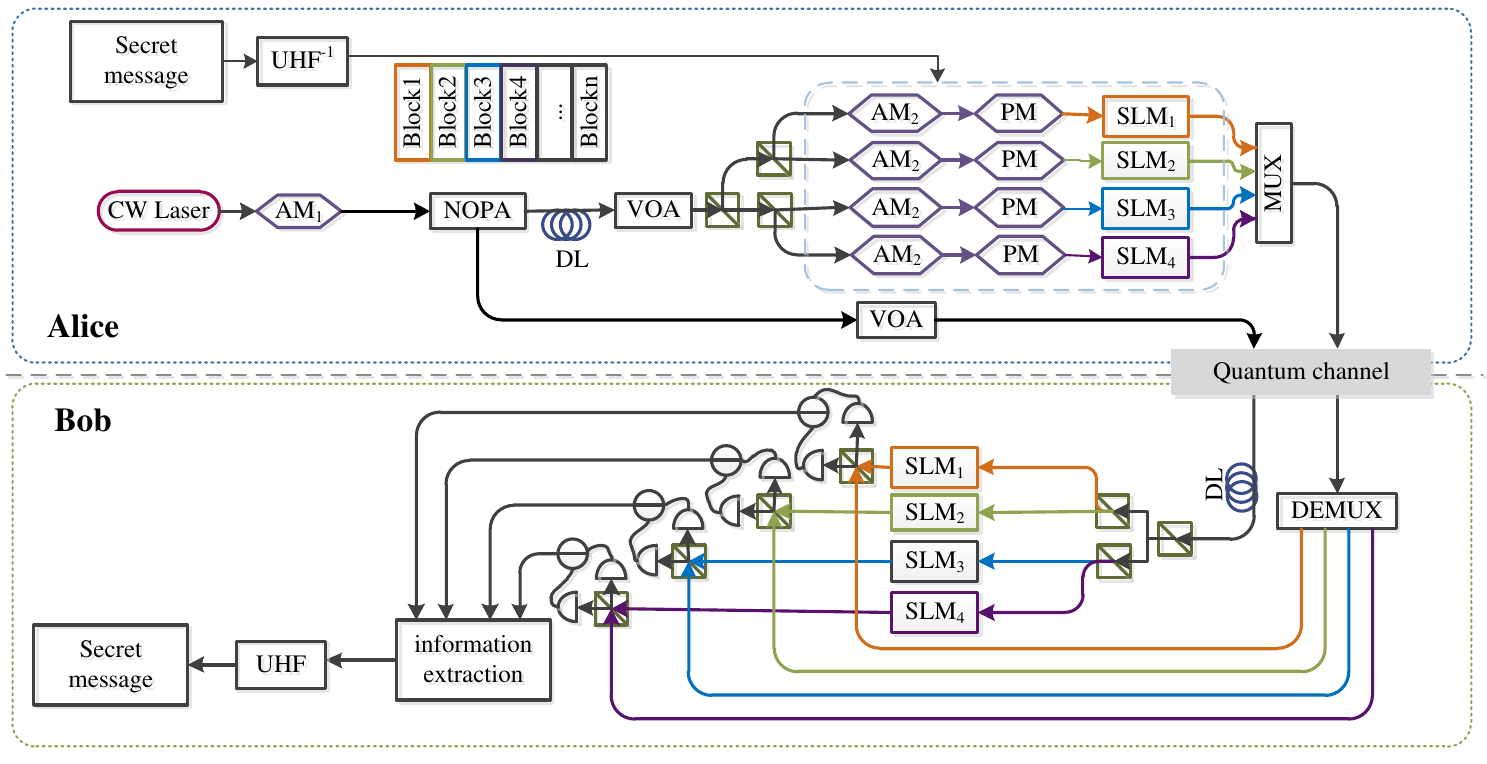}
   \caption{Experiment setup. $CW$, 1550nm continuous wave light 
   source; $ UHF$, hash forward transformation; $DL$, fiber optic delay 
   line; $AM$, amplitude modulator; $PM$, phase modulator; $VOA$, 
   variable optical attenuator; $BLOCKi$, the information block after 
   partitioning; $NOPA$, non degenerate optical parametric amplifier; 
   $SLMi$, spatial light modulator; $MUX$, OAM multiplexing; $DEMUX$, 
   OAM demultiplexing.}
   \label{fig4}
 \end{figure*}
\section{\label{sec:level3}Experimental analysis} 
 
 This section introduces the experimental analysis of the proposed scheme. First of all, the way of setting up the experimental platform is introduced. Then, the parameters of the experimental data are estimated, and the results finally are analyzed to draw a conclusion.

\subsection{\label{sec:level31}Experiment implementation}
This protocol requires the multiplexing of four OAM modes, and the experimental platform setup is shown in the Fig. \ref{fig4}. In this experiment, we divide the information block into four parts and estimate the parameters of the 800 blocks data results, and based on the measurement results, the core parameters of the system are estimated to evaluate the effectiveness of the experiment.
 
  Alice employs a commercial frequency-stabilized 1550 nm wavelength continuous-wave (CW) laser with 150 kHz linewidth as the signal laser, and uses a customized 10 GHz Lithium Niobate electro-optic 
  amplitude modulator to transform the CW light into an optical pulse train with 2 ns pulse width and at a repetition rate of 50 MHz. Then, Alice prepares two-mode squeezed entangled state by passing optical pulse through non-degenerate optical parametric amplifier, one as signal light and another as detection light. After that, the detection light is sent to Bob after passing variable optical attenuator \cite{r27}. After the signal light is adjusted through the optical fiber delay line, it is divided into four-way through a three-component beam splitter. Alice preprocesses the secret messages according to Sec.  \ref{sec:level21} and rotates the four modulated signal lights through spatial light modulators to four orthogonal OAM, and then send them to Bob after a multiplexer. 
  
  Bob receives two lights passing through 10 km quantum channel. Concretely speaking, when Bob receives the detection light, and then processes the detection light through the three-component beam splitter and spatial light modulators same as Alice for subsequent CV Bell detection. When Bob receives the modulated signal light, it is demultiplexed to combine with the detection light for CV Bell measurement. Finally, the messages after demultiplexing and measuring is extracted and the whole experimental process is completed.
  
  The correlation between the data of the four OAM modes and the overall data is shown in Fig. \ref{fig5}, and the estimated parameters are shown in Tab. \ref{tab1}. From Fig. \ref{fig5}, it can be seen that the correlation of each OAM channel tends to be consistent, consistent with the overall data correlation, and does not affect the overall correlation, which verifies the effectiveness and feasibility of the experiment. According to Tab. \ref{tab1}, the estimated results of the system parameters are consistent with the data correlation in Fig. \ref{fig5}, which verifies the feasibility of orbital angular momentum multiplexing in quantum communication. 
  
\subsection{\label{sec:level32}Parameter estimation}
 This section performs parameter estimation on the variable after passing through the channel to ensure the validity of experimental data, and to estimate the of information obtained by Alice and Bob with the help of calculating the error rate of this experiment. In parameter estimation, it is necessary to obtain the confidence interval. Here, we introduce the normal model \cite{r21,r22} of Alice and Bob's variables:
\begin{equation}
y=tx+z, 
\label{eq30}
\end{equation}
where $t=\sqrt{\eta T} \in R$. $z$ follows the normal distribution of variance $\sigma ^2=N_0+\eta T\varepsilon +v_{el}$. Similar to the analysis in \cite{r22}, the maximum likelihood estimator $\hat{t}$ , $\hat{\sigma } ^2$ and $\hat{\sigma } _0^2$   can be calculated 
according to the following equation:
\begin{equation}
\left\{
\begin{aligned}
&\hat{t} =\frac{\sum_{i=1}^{\mathcal{N} }x_iy_i}{\sum_{i=1}^{\mathcal{N} 
}x^2_i}, 
\hat{\sigma } ^2=\frac{1}{\mathcal{N} } \sum_{i=1}^{\mathcal{N} } 
(y_i-\hat{t}x_i )^2, 
\cr
&\hat{\sigma } _0^2=\frac{1}{\mathcal{N} '} \sum_{i=1}^{\mathcal{N} '} 
y_{0i}^2, 
\hat{V} _a=\frac{1}{\mathcal{N} } \sum_{i=1}^{\mathcal{N} }x_i^2 . 
\label{eq31}
\end{aligned}
\right.
\end{equation}

And these estimators are independent and satisfy the following 
distribution:
\begin{equation}
\left\{
\begin{aligned}
\label{eq32}
&\hat{t} \sim N\left ( t,\frac{\sigma ^2}{\sum_{N}^{i=1}x_i^2 }  \right 
)  , \cr
&\frac{\mathcal{N} \hat{\sigma }^2  }{\sigma ^2}, \frac{\mathcal{N}' 
\hat{\sigma }_0^2  }{\sigma_0 ^2}, \frac{\mathcal{N} \hat{V_a} 
}{V_a}\sim \chi ^2(m-1)   . 
\end{aligned}
\right.
\end{equation}

\begin{figure*}[!htb]\center
  \centering
  \subfigure[\;{OAM Channel 1-4}]{\label{fig:4a}
  \includegraphics[width=8.4cm]{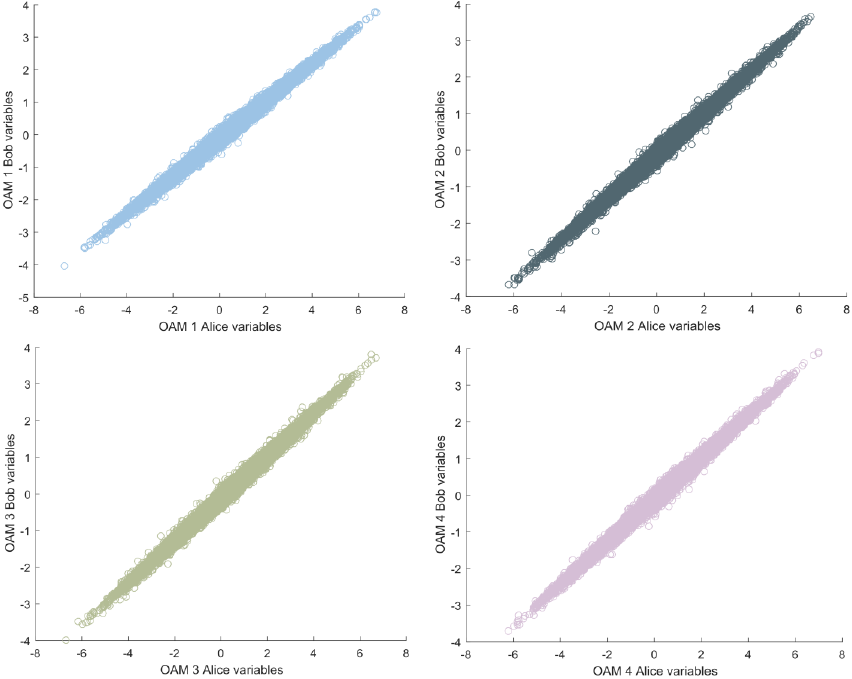}}
  \subfigure[\;{Merge and combine}]{\label{fig:4}
  \includegraphics[width=7.2cm]{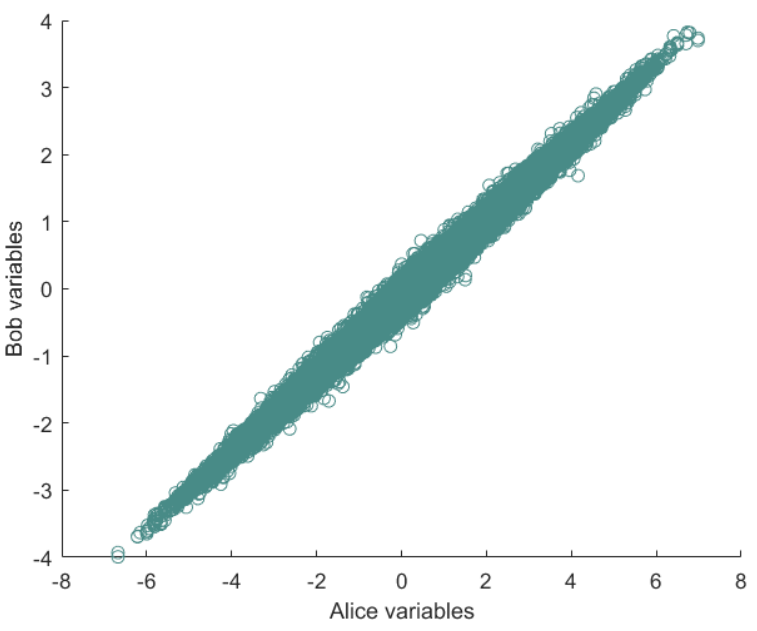}}

  \caption{Autocorrelation analysis results of four channel 
   multiplexing OAM data. Figs. (a) show the cross-correlation 
   analysis results of each OAM channel, and Figs. (b) shows the 
   cross-correlation analysis results of all data obtained after 
   merging the four channel data.}
  \label{fig5}
\end{figure*} 
 
From this, the confidence intervals of these parameters can be calculated:
\begin{equation}
\left\{
\begin{aligned}
&t\in \left [ \hat{t}-\triangle t,\hat{t}+\triangle t   \right ]  , 
\cr
&\sigma ^2\in \left [ \hat{\sigma }^2-\triangle \sigma ^2,\hat{\sigma 
}^2+\triangle \sigma ^2   \right ]  , 
\cr
&\sigma _0^2\in \left [ \hat{\sigma }_0^2-\triangle \sigma_0 
^2,\hat{\sigma }_0^2+\triangle \sigma _0^2   \right ]   , 
\cr
&V _a\in \left [ \hat{V }_a-\triangle V_a,\hat{V }_a^2+\triangle V _a   
\right ]   , 
\label{eq33}
\end{aligned}
\right.
\end{equation}

\begin{widetext}
	\begin{ruledtabular}
		\begin{table*}[htbp]
			\caption{Parameter estimation results of experimental variable. }
			\label{tab1}
			\centering
			\begin{tabular}{ c c c c c c }				\multicolumn{1}{c}{\multirow{1}{*}{Parameter$\setminus$Channels}} & \multirow{1}{*}{\shortstack{OAM 1}} &\multirow{1}{*}{\shortstack{OAM 2}} &\multirow{1}{*}{\shortstack{OAM 3}} &\multirow{1}{*}{\shortstack{OAM 4}} &\multirow{1}{*}{\shortstack{MUX channel}}\\				
                \hline				
				\multirow{1}{*}{$\hat{T}$}      &$0.6273$ &$0.6284$ &$0.6271$ &$0.6274$ &$0.6275$ \\				                             				
				\multicolumn{1}{c}{$\hat{\varepsilon }(SNU) $}      &0.0184&0.0184&0.0184&0.0184&0.0184 \\			
\end{tabular}
		\end{table*}
	\end{ruledtabular}
\end{widetext}
where,
\begin{equation}
\left\{
\begin{aligned}
&\triangle t=z_{\varepsilon _\text{{PE}}/2}\sqrt{\frac{ \hat{\sigma}^2 }{mV_a} } ,\triangle \sigma ^2=z_{\varepsilon _\text{{PE}}/2}\frac{\hat{\sigma}^2 \sqrt{2}  }{\sqrt{m} }, \cr
&\triangle V_A=z_{\varepsilon _\text{{PE}}/2 }\frac{\hat{V_A}^2\sqrt{2} }{\sqrt{N}} ,\triangle \sigma _0^2=z_{\varepsilon _\text{{PE}}/2 }\frac{\hat{\sigma _0}^2\sqrt{2} }{\sqrt{N'}.} 
\label{eq333}
\end{aligned}
\right.
\end{equation}
wherein, parameter $z_{\varepsilon _\text{{PE}}/2}$ satisfies the following relationship,
\begin{equation}
\left\{
\begin{aligned}
&1-\text{erf}\left (\frac{ z_{\varepsilon _\text{PE}/2}}{\sqrt{2} }  \right ) =\frac{\varepsilon _\text{PE}}{2} \cr
&-\text{erf}\left ( x \right ) =\frac{2}{\sqrt{\pi } } \int_{0}^{x} \exp\left ( -\tau ^2 \right ) d\tau 
\end{aligned}
\right.
\end{equation}
%where, the interval limit can be calculated based on the estimate, as follows:
%\begin{equation}
%\left\{
%\begin{aligned}
%&\hat{T}=\frac{\hat{t}^2 }{\eta }  , 
%\cr
%&\hat{\epsilon} =\frac{\hat{\sigma} ^2-\hat{\sigma _0}^2 }{\hat{t}^2 }  
%.
%\label{eq34} 
%\end{aligned}
%\right.
%\end{equation}

\begin{figure}
   \centering
   \includegraphics[scale=0.48]{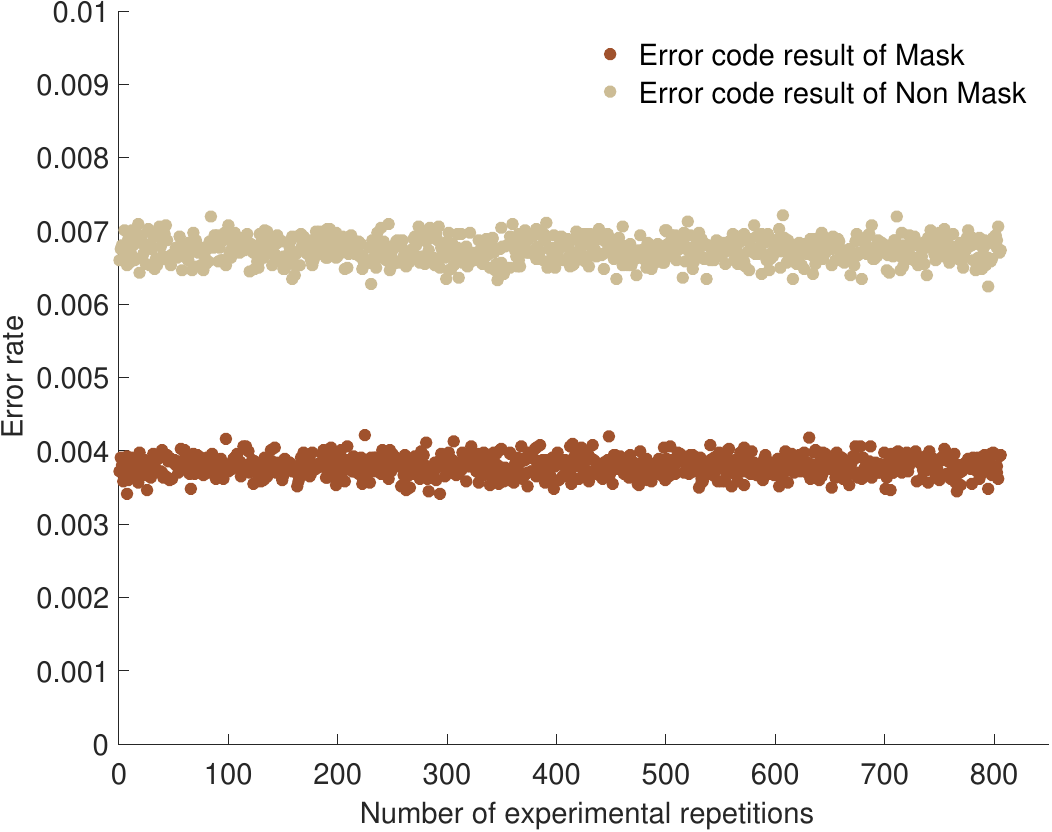}
   \caption{Error rate analysis. The figure shows the estimated bit 
   error rate of more than 800 groups of different secret information 
   transmitted by the proposed protocol, and the light colors represent results obtained without masking operations under the same conditions, in which the efficiency 
   of the detector is 0.5.}
   \label{fig6}
 \end{figure}
The system error rate results in masked messages extraction are shown in Fig. \ref{fig6}. It can be seen that the error rate of the experimental system remains stable, with an average of 0.0038. This verifies that mask coding can be used to protect the transmission security of information in the channel and resist the influence of channel noise, further reflecting the stability and feasibility of the experimental system.
 
 \begin{figure}
   \centering
   \includegraphics[scale=0.65]{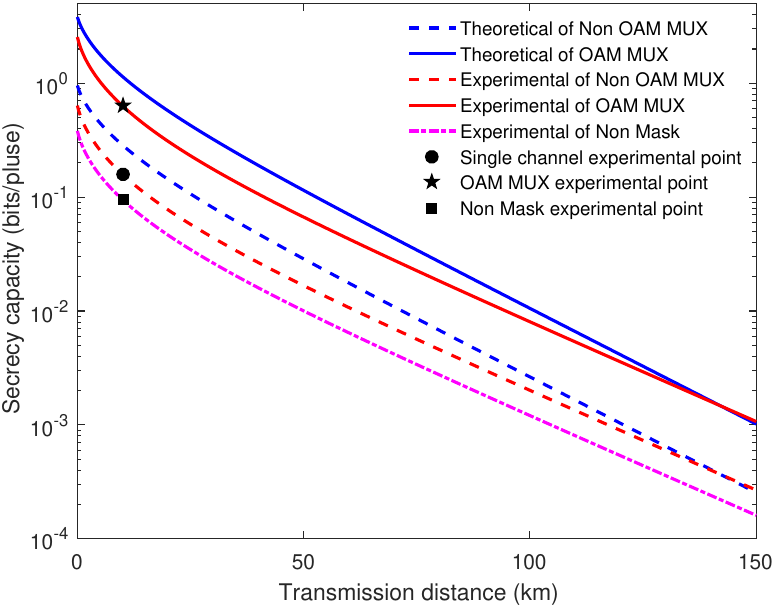}
   \caption{The blue line indicates the amount of mutual information 
   between Alice and Bob, and the red line indicates the amount of 
   mutual information with eavesdroppers; the dotted line indicates the 
   results without OAM multiplexing, and the solid line indicates the 
   results after OAM multiplexing; 
   the dotted dashed line indicates the results without OAM multiplexing and masking;  
   the $``\star"$ mark refers to the 
   results obtained through experiments, and the $``\bullet"$ mark 
   refers to the results obtained without reuse under the same 
   conditions, the $``\blacksquare"$ mark 
   refers to the results obtained without reuse and mask under the same 
   conditions.}
   \label{fig7}
 \end{figure}  
 Based on the parameter estimation results in Fig. \ref{fig5} and Tab. \ref{tab1}, as well as the transmission error rate in Fig. \ref{fig6}, the relationship between system transmission distance and secrecy capacity is shown in the Fig. \ref{fig7}. The four curves in Fig. \ref{fig7} represent the theoretical value of multiplexing, experimental value of multiplexing, single channel theoretical value, and single channel experimental value from top to bottom, respectively. At the same distance, this scheme has a significant improvement in secrecy capacity compared to existing schemes, which is determined by the number of observable orbital angular momentum in the experiment. At the same time, there is a certain gap between the experimental and the theoretical value, which is caused by channel errors. In the future, further research can be conducted on quantum state oriented coding and decoding techniques to reduce transmission errors and further improve system performance.
 
\section{\label{sec:level4}Conclusion and discussion} 
Based on the QSDC scheme from the perspective of protocol and experiment. On the premise of ensuring the security and performance of the protocol, the combination of mask-coding technology and orbital angular momentum multiplexing not only improves the security capacity of the system, but also can effectively combat the impact of channel noise and Eve on communication. The experimental results show that in the transmission of $800$ information blocks $\times  1310 $ bits length, the statistical average of bit error rate for this scheme is $0.38\%$, the system excess noise is $0.0184$ SNU, and the final security capacity is $6.319\times10^{6}$ bps. Therefore, the proposed protocol can ensure that the legitimate parties can obtain more confidential information and improve security and transmission efficiency. The research on the actual security of the protocol will further optimize our protocol and deal with more complex communication environment, which will be the focus of our work in the future. Therefore, the proposed protocol will further enrich the QSDC technology and promote the implementation of a practical QSDC system.

\begin{acknowledgments}This work is supported by the National Natural Science Foundation of China (62071381, 62301430), Shaanxi Fundamental Science Research Project for Mathematics and Physics (23JSY014), Scientific research plan project of Shaanxi Education Department (Natural Science Special Project (23JK0680)), Young Talent Fund of Xi'an Association for Science and Technology (959202313011) .
\end{acknowledgments}

%\section*{References} 

\end{document}